\documentclass[twocolumn]{jpsj2} %% two-column layout
\usepackage{multicol}

\title{The Kondo Effect in Non-Equilibrium Quantum Dots:\\ Perturbative Renormalization Group}

\author{\textsc{A. Rosch}$^{1}$,
  \textsc{J. Paaske}$^{2}$,  \textsc{J. Kroha}$^{3}$ and
 \textsc{P. W\"olfle}$^{2}$}

\inst{$^{1}$Institut f\"ur Theoretische Physik, Universit\"at zu K\"oln, 50937 K\"oln, Germany \\
$^{2}$Institut f\"ur Theorie der Kondensierten Materie, Universit\"at Karlsruhe, 76128 Karlsruhe, Germany \\
$^{3}$Physikalisches Institut, Universit\"at Bonn, 53115 Bonn,
Germany}

\abst{ While the properties of the Kondo model in equilibrium are
very well understood,  much less is known for Kondo systems out of
equilibrium. We study the properties of a quantum dot in the Kondo
regime, when a large bias voltage $V$ and/or a large magnetic field
$B$ is applied. Using the perturbative renormalization group
generalized to stationary nonequilibrium situations, we calculate
renormalized couplings, keeping their important energy dependence.
We show that in a magnetic field the spin occupation of the quantum
dot is non-thermal, being controlled by $V$ and $B$ in a complex way
to be calculated by solving a quantum Boltzmann equation. We find
that the well-known suppression of the Kondo effect at finite $V \gg
T_K$ (Kondo temperature) is caused by inelastic dephasing processes
induced by the current through the dot. We calculate the
corresponding decoherence rate, which serves to cut off the RG flow
usually well inside the perturbative regime (with possible
exceptions). As a consequence, the differential conductance, the
local magnetization, the spin relaxation rates and the local
spectral function may be calculated for large $V,B \gg T_K$ in a
controlled way.
 }

\kword{Kondo effect, quantum dots, non-equilibrium, decoherence,
perturbative renormalization group}

\newcommand{\w}{\omega}
\newcommand{\sign}{\text{sign}}
\newcommand{\s}{\sigma}

\newcommand{\g}{\gamma}
\newcommand{\glr}{g_{LR}}
\newcommand{\gll}{g_{LL}}
\newcommand{\gd}{g_{d}}
\newcommand{\grr}{g_{RR}}

\newcommand{\Nf}{N_0}
\newcommand{\be}{\begin{equation}}
\newcommand{\bea}{\begin{eqnarray}}
\newcommand{\ee}{\end{equation}}
\newcommand{\eea}{\end{eqnarray}}
\newcommand{\bwt}{\begin{widetext}}
\newcommand{\ewt}{\end{widetext}}
\renewcommand{\vec}[1]{{\mathbf #1}}

\begin{document}
\maketitle

\section{Introduction}
After 40 years of research, the single-impurity Kondo
model\cite{Kondo64} and its variants are certainly among the best
understood strongly correlated systems. An amazingly broad spectrum
of powerful theoretical methods\cite{Hewson93} have been developed
to describe the physics of a magnetic impurity in a metal, ranging
from exact analytical solutions using the Bethe Ansatz, bosonization
and conformal field theory, diagrammatic methods, flow equations,
perturbative renormalization group to sophisticated numerical
techniques like the numerical renormalization group. Using these
tools an almost complete understanding of thermodynamic, transport
and spectral properties of the single-impurity Kondo model has been
achieved \cite{Hewson93}.

There is, however, one important field about which in comparison
little is known: Kondo physics out of equilibrium, a question of
both experimental and theoretical importance. Non-equilibrium can,
for example, be reached when a finite current is driven through a
Kondo system (see below).

The study of the Kondo effect out of equilibrium started both
experimentally and theoretically
 in the late 1960's when the tunneling between two metals through
insulating barriers was
investigated\cite{Shen68,Nielsen70,Appelbaum72,Wallis74,Wolf70,Bermon78}
as a function of bias voltage.  The observed logarithmic {\em enhancement} of
the tunneling for low temperature and small voltages was attributed
to exchange tunneling through magnetic impurities within the
barriers by Appelbaum \cite{Appelbaum66,Appelbaum67a} and
Anderson\cite{Anderson66}.
The relevant impurities were argued\cite{Anderson66} to reside close to
one side of the junction, and the localized spin was therefore tacitly
assumed to be in equilibrium with the metal closeby. The tunneling to the
other metal is weak and the conductance in such a
situation is determined by the equilibrium density of states on the
impurity\cite{Solyom68,Appelbaum70}.  Subsequent work by
Ivezi\'{c}\cite{Ivezic75} considered also impurities deep inside the
junction, which were pointed out to require a full non-equilibrium
treatment. A good review of these earlier works may be found in
Ref.~\cite{Wolf85}.

With the progress in nanotechnology, it became possible to realize
Kondo physics in quantum
dots\cite{Goldhaber98,Cronenwett98,Schmid98,Nygaard00,vanderWiel00,Goldhaber98b}.
A dot in the Coulomb blockade regime which carries a net spin can be
mapped to the Kondo model\cite{Glazman88,Ng88}. The resonant
tunneling through the dot leads to a removal of the Coulomb
blockade, i.e. an increase of the conductance from small values up
to the quantum limit, as the Kondo resonance develops.  A finite
bias voltage $V$ drives the system out of equilibrium. It has been
observed that the Kondo effect is quenched by raising the transport
bias voltage V well above the Kondo temperature $T_K$, i.e. $V \gg
T_K$, and that the presence of a magnetic field splits the zero-bias
conductance peak into two distinct peaks, located at bias voltages
roughly equal to plus and minus the Zeeman splitting of the spin on
the
dot\cite{Goldhaber98,Cronenwett98,Schmid98,Nygaard00,vanderWiel00}.
Similar experiments were also possible in metallic nanoconstrictions
where it is possible to measure transport through a single magnetic
impurity \cite{Ralph94}, however, in contrast to quantum dots it is
not possible to control system parameters in such devices.

The most straightforward way to study quantum dots out of
equilibrium is to apply a finite dc bias voltage and to measure the
current-voltage characteristics. There have, however, been a few
remarkable experiments which go further. For example, Franceschi
{\it et al.} \cite{noneqDistr} managed to measure the splitting of
the Kondo resonance by a dc bias voltage by using a three-lead
configuration. Another set of questions can be addressed by studying
the response to time-dependent external fields induced, for example,
by external irradiation by a microwave field with frequency $\omega$
\cite{Goldin98,Kaminski99,hettler95,ng96,lopez}. Kogan {\it et al.}
\cite{kogan} recently succeeded in observing satellites of the Kondo
effect separated by $\omega$ in such an experiment.

Kondo impurities  also have a pronounced effect on the distribution
of electrons in mesoscopic wires, which are driven out of
equilibrium by an applied bias
 voltage\cite{pothier.97,anthore.03,schopfer.03}. The inelastic
scattering from the magnetic impurities at finite bias leads to a
characteristic broadening\cite{glazman.01,goeppert.01,kroha.02} of
the electronic distribution functions which can be measured in
tunneling experiments\cite{pothier.97,anthore.03,schopfer.03}.

Theoretically, the Kondo effect out of equilibrium has been studied
by a number of methods ranging from perturbation theory
\cite{Appelbaum66,Appelbaum67a,Appelbaum67b,Solyom68,Kaminski99,Parcollet02,Rosch03a,paaske1,paaske2,Sivan96,hersh91,Fujii,Oguri},
equations of motions and self-consistent diagrammatic methods
\cite{Meir93,Wingreen94,Hettler94,ng96,Konig96,Norlander99,Plihal00,schiller00,Krawiec02,Rosch01}
(using the so-called non-crossing approximation), slave-boson
mean-field theories \cite{aguado,slaveColeman,han}, exact solutions
for some variants of the Kondo model with appropriately chosen
coupling constants \cite{Schiller95}, the construction of
approximate scattering states starting from Bethe ansatz
solutions\cite{Konik01},
 to perturbative renormalization group
\cite{Kaminski99,Rosch03a,spectral}(reviewed below). It is, however,
important to note, that many of the methods which have been so
successful in equilibrium cannot or have not yet been generalized
even to the simplest steady-state non-equilibrium situation. One of
the reasons for this is that the current-carrying state at finite
bias is a highly excited many-body state of the system.
   Therefore, all methods which by construction focus on ground
   state properties are not readily generalized to such a situation.

In the following sections, we will review our version
\cite{Rosch03a} of  perturbative renormalization group (RG) in the
presence of a finite bias voltage -- formulated in the spirit of
Anderson's poor man's scaling \cite{Anderson70}. We generalize the
RG equations presented in Ref.~\cite{Rosch03a} for a symmetric dot
to an arbitrary exchange coupling of the spin to the conduction
electrons of two attached leads with different electrochemical
potential. The central goal is to perform a controlled calculation
in the limit of either large bias voltage $V\gg T_K$ or large
magnetic field   $B \gg T_K$ or large probing frequency $\w \gg
T_K$. We suggest a method to calculate the behavior of the
conductance and other physical quantities like magnetization or
spectral function in leading order of the small parameter
$1/\ln[\max(V,B,\w)/T_K]$. We will argue that the perturbative
renormalization group differs in three main aspects from its
counterpart in equilibrium \cite{Anderson70}: First, the
magnetization of the Kondo impurity (or in general the occupation
probabilities of the quantum states) has to be calculated
self-consistently  from appropriate quantum Boltzmann equations
\cite{Parcollet02,Rosch03a,paaske1}. This leads to an unusual
dependence e.g. of the spin susceptibility on $V$ and to a novel
structure of the logarithmic corrections. Second, the perturbative
renormalization group has to be formulated in terms of
frequency-dependent coupling {\em functions} instead of coupling
{\em constants}. The reason is that electrons in an energy {\em
window} set approximately by the external bias voltage contribute to
the low-energy properties (see Fig.~\ref{window}) and their position
within this window will affect their effective coupling to the spin.
Therefore, the effective renormalized coupling of the conduction
electrons to the local spin will depend explicitly on their energy.
Third, decoherence effects are much more important out of
equilibrium \cite{Meir94,Kaminski99,Rosch01,Rosch03a,paaske2}. A
finite current will induce noise and thereby induce dephasing of the
coherent spin-flip processes responsible for the Kondo effect. We
will show that it is precisely due to those dephasing effects that a
controlled calculation at large voltages is possible.

\section{Perturbation Theory}

We consider the Kondo  Hamiltonian
\begin{multline} \label{H}
 H=\sum_{\alpha=L,R,{\bf k},\s}(\varepsilon_{{\bf k}}-\mu_{\alpha})
c^{\dagger}_{\alpha{\bf k}\s}c_{\alpha{\bf k}\s}- B S_{z}\\
 +\frac{1}{2} \!\!\!\sum_{\alpha,\alpha'=L,R,{\bf k},{\bf k}',\s,\s'}
\!\!\!\!\!J_{\alpha'\alpha}\, \vec{S}\cdot (c^{\dagger}_{\alpha'{\bf
k}'\s'} {\boldsymbol \tau}_{\s'\s}c_{\alpha{\bf k}\s}),
\end{multline}
where $\mu_{L,R}=\pm V/2$. $\vec{S}$ is the spin 1/2 on the dot and
${\boldsymbol \tau}$ are the Pauli matrices. This model describes a
quantum dot coupled to two leads (the left and right electrons
described by  $c^\dagger_{L/R,{\bf k}\s}$) in which the number of
electrons on the dot is fixed to an odd integer by an interplay of
gate voltage and charging energy. In this Coulomb blockade regime,
effectively a single spin is localized on the dot which interacts
via an exchange coupling $J_{\alpha \alpha'}$ arising from tunneling
processes from and to the leads involving virtual excitation of the
dot. In cases where the quantum dot can be described by a simple
Anderson model, one obtains $J_{LR}=J_{RL}=\sqrt{J_{LL} J_{RR}}$. In
more complex situations, $J_{\alpha \alpha'}$ can be an arbitrary
symmetric $2\times 2$ matrix. In general, a further cotunneling term
$V c^\dagger_\alpha c_{\alpha'}$ exists. As such a term does not
flow to strong coupling under renormalization group, it can be
safely neglected (furthermore, such a term vanishes for an
appropriate choice of a gate voltage in the middle of the Coulomb
blockade valley). We shall use the dimensionless coupling constants
$g_{\alpha \alpha'}=\Nf J_{\alpha \alpha'}$, where $\Nf$ is the
local density of states in the leads. For simplicity, we have
coupled in Eq.~(\ref{H}) the magnetic field (measured in units where
$g \mu_B=1$) only to the local spin. An extra coupling to the
electrons would effectively lead only to a small renormalization of
the $g$ factor. We assume a constant density of states in
the leads in a frequency range of order $B$ and $V$. In such a
situation the density of states at the Fermi level remains
unmodified in the presence of $B$.

In order to be able to use standard diagrammatic techniques, we
represent the local spin $\vec{S}=\frac{1}{2}\sum_{\gamma
\gamma'}f^{\dagger}_{\gamma}
{\boldsymbol\tau}_{\gamma\gamma'}f_{\gamma'}$ by pseudo-fermions  in
the sector of the Hilbert space with $\sum_{\gamma}
f^{\dagger}_{\gamma}f_{\gamma}=1$. Details of the Keldysh
diagrammatic method \cite{Rammer86} used throughout this paper and
how the exact projection to the physical Hilbert space of the pseudo
Fermions is performed  can be found in Ref.~\cite{paaske1} where the
perturbation theory for the Kondo model is derived in detail to
leading logarithmic order.

Before setting up a renormalization group scheme, it is useful to
investigate the results of  perturbation theory, e.g. for the spin
susceptibility. Out of equilibrium, the calculation of the
susceptibility has to be done with some care, the bare perturbation
theory diverges even at finite temperature $T$. Physicswise this
arises because for $J=0$ the magnetization of the completely
uncoupled dot is undetermined. While in equilibrium any
infinitesimal coupling leads to the usual thermal occupation {\em
independent} of all details of the coupling, this is not the case
out of equilibrium, where the details of the coupling do matter.
Practically this implies that one has to calculate the occupation
function even in the limit of vanishing $J$ from some type of
(quantum-) Boltzmann equation. In diagrammatics, this is given by
one component of a (self-consistent) Dyson equation in Keldysh space
(see Refs.~\cite{Parcollet02,paaske1} for details). To lowest order
this is just the well-known Boltzmann equation with golden-rule
transition rates. To obtain the steady state, the spin-flip rate
from $\uparrow$ to $\downarrow$  has  to equal the rate in the
opposite direction and therefore
\begin{multline}
n_{\uparrow}\sum_{\alpha,\alpha{'} = L,R} g_{\alpha \alpha'}^2 \int
d\omega \, f_{\omega
- \mu_\alpha} \Big(1-f_{\omega-\mu_{\alpha{'}}-B}\Big) = \\
 = n_{\downarrow} \sum_{\alpha,\alpha{'}=L,R} g_{\alpha \alpha'}^2\int d\omega \,
f_{\omega - \mu_\alpha}\Big(1 - f_{\omega - \mu_{\alpha{'}}+B}\Big)\
. \label{boltz}
\end{multline}
where $f_\w$ is the Fermi function. Solving for $n_\uparrow
-n_\downarrow$ in the limit $B\to 0$, one finds
\cite{Parcollet02,Rosch03a,paaske1}
\begin{eqnarray}
\chi&=& \frac{2\glr^{2}+\gll^2+\grr^2}
{2V\coth\left(\frac{V}{2T}\right)\glr^{2}+2 T (\gll^2+\grr^2) }.
\label{chi0}
\end{eqnarray}
In equilibrium, $V=0$, all coupling constants cancel out and one
obtains $\chi=1/(2 T)$ to zeroth order in the coupling constants.
For a finite voltage, however, the result depends on the ratio of
the coupling constants even for infinitesimal $J_{\alpha \alpha'}$.
For large voltages one finds $\chi\sim 1/V$.

One order higher, the logarithmic correction characteristic for the
Kondo effect\cite{Kondo64} arises and one obtains
\cite{Rosch03a,paaske1}, with $\gd=(\gll+\grr)/2$,
%\onecolumn
%\begin{widetext}
%\bwt
%\begin{multicols}{1}
\begin{eqnarray}\hspace{15mm}\chi=\frac{
2\glr^{2}\left(1+2\gd\ln\frac{D}{T}+2\gd\ln\frac{D}{|V|}\right)
+\sum_{\g}g_{\g\g}\left(g_{\g\g}+2g_{\g\g}^2\ln\frac{D}{T}+
                                   2\glr^{2}\ln\frac{D}{|V|}\right)
} {
2V\coth\left(\frac{V}{2T}\right)\glr^{2}\left(1+4\gd\ln\frac{D}{|V|}\right)
+2 T\sum_{\g}g_{\g\g}\left(g_{\g\g}+2g_{\g\g}^2\ln\frac{D}{T}+
                                      2\glr^{2}\ln\frac{D}{|V|}\right)
}\hspace{5mm}\label{eq:chi}
\end{eqnarray}
%\end{multicols}{1}
%\end{widetext}
%\ewt%\twocolumn
where the bandwidth $D$ cuts off the logarithmic singularities at
high energies. Note that logarithmic corrections arise already to
{\em linear} order in $g$ with prefactors which again depend on the
ratios of the coupling constants. In the limit of $T\gg V$, in
contrast, the logarithms all take the form $\ln(D/T)$, the
corrections in the numerator and in the denominator cancel, and we
are left with the usual Curie law $\chi=1/2T$, as in equilibrium the
first logarithmic correction arises only to order $g^2$, not
included in Eq.~(\ref{eq:chi}). This shows again that in the
presence of a finite bias voltage, the structure of perturbation
theory is strongly modified.

\section{Perturbative Renormalization Group}

\begin{figure}
\includegraphics[width=  0.95 \linewidth]{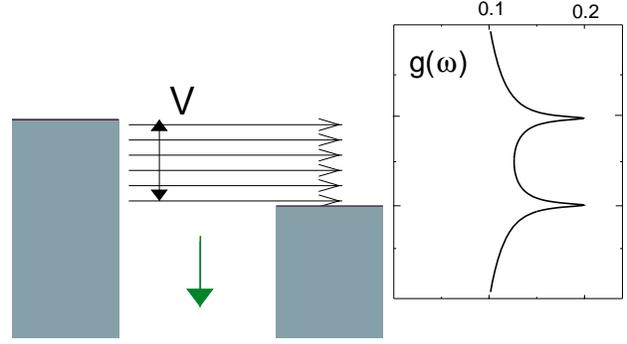}
\caption{\label{window} Due to the finite bias voltage, the
electrochemical potential of the right and left lead differ by $V$.
As all electrons in an energy window of width $V$ contribute to low
energy properties, it is necessary to take into account that the
effective exchange coupling  of the electrons depends on their
energy. Right panel: frequency dependent effective coupling $g(\w)$
calculated from Eq.~(\ref{gdw}) for $V/T_K=100$.}
\end{figure}

Even for small coupling constants $g$ and for sufficiently large $V$
and $B$, such that $V,B \gg T_K$, but still in the scaling regime
$V,B \ll D$, such that $g \ln  D/(V,B) < 1$, bare perturbation
theory converges slowly.  It is necessary to sum the leading
logarithmic contributions in all orders of perturbation theory. Only
after such a resummation another important property of the Kondo
model becomes manifest: its physics depends only on a single energy
scale, the Kondo temperature $T_K$. At least for sufficiently small
$T_K$, all microscopic complications, like band-structure effects,
the energy dependence of the couplings, etc.  can be absorbed in the
value of $T_K$ but will otherwise not affect low-energy properties
which are completely {\em universal} functions of $T/T_K$ or $B/T_K$
for $T,B \ll D$.
 In technical terms, this arises because the Kondo model is renormalizable. A similar behavior is
expected in the presence of a bias voltage $V$. We expect e.g. that
the conductance $G$ through a quantum dot in the Kondo regime will
be a universal function, $G(V/T_K,B/T_K,T/T_K)$ at least for
symmetric coupling $J_{LL}=J_{LR}=J_{RR}=J$. (Note, however, that
the conductance and all other physical observables will
\vspace*{2.3cm} depend on the ratios of the coupling constants for
$V>0$ as is obvious from the simple fact that no current is flowing
for $J_{LR}\to 0$.)

Anderson\cite{Anderson70} pioneered with his ''poor man's scaling''
a powerful method to resum the leading logarithmic corrections of
the Kondo model in a controlled way: perturbative renormalization
group (RG). This approach makes use of the fundamental idea that a
small change of the cut-off $D$ can be  absorbed into a redefinition
of the coupling constants $g$.   As long as the cutoff-dependent
running coupling constant $g(D)$ is small, the change of $g$ under
an infinitesimal change of $D$, $\partial g/\partial \ln D$, may be
calculated in perturbation theory. In the equilibrium Kondo problem
for vanishing magnetic field and temperature, the coupling constant
grows when the cutoff is reduced,  leaving the perturbative regime
$g < 1$ when the running cutoff $D$ is of the order of $T_K$. The
flow of the coupling constant to infinity leads finally to a
complete screening of the localized spin.

In the presence of a bias voltage $V \gg T_K$ the situation is more
complex as we will show below: while the RG flow is {\em not} cutoff
by the voltage itself, it is stopped\cite{Rosch01,Rosch03a} before
the strong coupling regime is reached by inelastic processes induced
by the finite current through the system (at least for the
experimentally most relevant case $J_{LR} = J_{LL}=J_{RR}$).

To derive renormalization group equations in such a situation, one
can for example start from a straightforward  generalization of
functional renormalization group methods, described e.g. in
Ref.~\cite{Salmhofer01}, to Keldysh diagrams. The basic idea is to
track the evolution of all one-particle irreducible diagrams when an
infrared cutoff is lowered. The advantage of this formulation is
that it naturally includes not only frequency dependent vertices but
also self-energy effects which are necessary to describe
decoherence. We will actually not follow this route here, but use a
simpler approach more in the spirit of Anderson's poor man's scaling
by investigating more directly the scaling properties of diagrams.
At each step we will try to simplify the equations as much as
possible: each approximation is tailored to be exact in leading
order of the small expansion parameter $1/\ln[\max(V,B,\w)/T_K]$.
 A different and considerably more involved real-time RG scheme
has been developed by Schoeller and K\"onig \cite{SK} but has not
yet been applied to this problem.

We start by analyzing the perturbation theory of various physical
quantities like the susceptibility (\ref{eq:chi}) and ask the question
whether a change in the cutoff $D$ can be absorbed in a redefinition
of the coupling constants. This turns out {\em not} to be  possible:
in Eq.~(\ref{eq:chi}) logarithmic corrections e.g. to $g_{LR}$ appear
as $2 \ln D/V$ in the denominator, but as $\ln D/V + \ln D/T$ in the
numerator.  When the cutoff $D$ gets smaller than $V$, different
physical quantities (or numerator and denominator of the same physical
quantity) seem to require different renormalizations of coupling
constants. This apparent contradiction to the principles of
renormalization group is easily resolved by realizing that one should
expect that the coupling constants depend on energy as explained above
and in Fig.~\ref{window}. For example, when analyzing the Boltzmann
equation (\ref{boltz}), one recognizes that the numerator in
(\ref{chi0}) arises from integrals which are confined to the vicinity
of the Fermi energy, whereas in the denominator the energy integral
covers a finite range of width $V$, and the couplings turn out to be
quite different in the two cases [c.f. Fig.~\ref{window}]. We
therefore investigate in the following the origin of the frequency
dependence of the vertex corrections.

\begin{figure}
$\displaystyle \frac{\partial}{\partial \ln D}
$\hspace{-0.2cm}\begin{minipage}{2cm}\includegraphics[height=1.8cm]{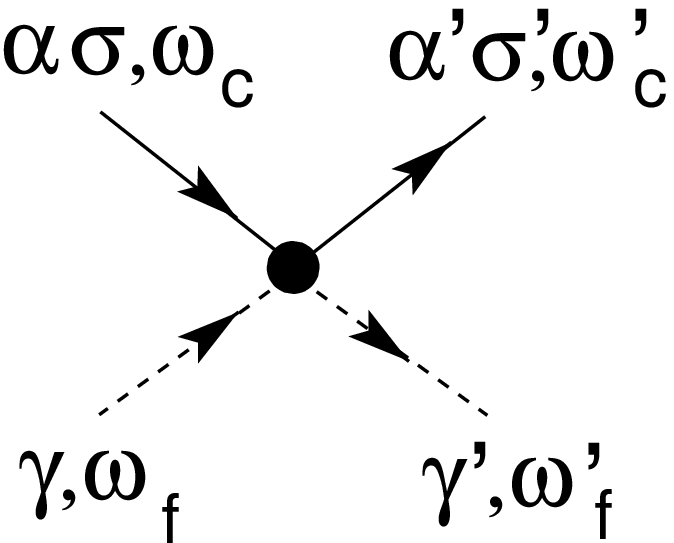}
\end{minipage}
\begin{minipage}{5.5cm}
\includegraphics[width=\linewidth]{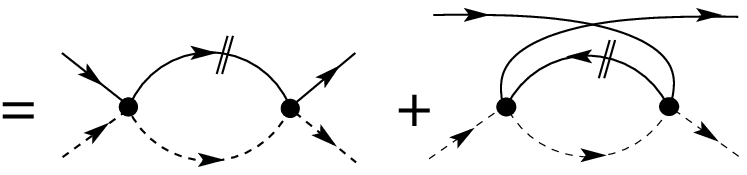}
\end{minipage}
\caption{\label{dia} Diagrammatic form of the RG equation
(\ref{fullRG}). The strokes symbolize derivatives with respect to
$\ln D$ and $\alpha,\alpha'=L/R$,
$\sigma,\sigma'=\uparrow\!\!\!/\!\!\downarrow$ and
$\gamma,\gamma'=\uparrow\!\!/\!\!\downarrow$ denote the quantum
numbers of the incoming and outgoing conduction electrons and pseudo
fermions, respectively. Their frequencies are given by
$\w_c,\w_c',\w_f,\w_f'$ with $\w_c+\w_f=\w_c'+\w_f'$.}
\end{figure}

All leading logarithmic terms in perturbation theory stem from the
simple vertex renormalizations shown in Fig.~\ref{dia} when the real
part $\approx 1/(\w \pm B/2)$ of the pseudo fermion Green function
is convoluted with the Keldysh component of the electron line $ - 2
\pi i \Nf \tanh[(\w-\mu_\alpha)/2T]$. Using cutoffs symmetric with
respect to $\mu_{L,R}$, respectively, one obtains at $T=0$
\begin{eqnarray}\label{logDer}
\frac{\partial}{\partial \ln D} \int_{-D}^D d\w \frac{\sign \
\w}{\w-\Delta \w}\approx 2 \Theta(D-|\Delta \w|)
\end{eqnarray}
where $\Delta \w$ depends on $V$, $B$ and the incoming and outgoing
frequencies. Approximating the logarithmic derivative by a Heaviside
step  function is valid in the two limits $|\Delta \w| \gg D$ and
$|\Delta \w| \ll D$. To leading order in our small parameter
$1/\ln[\max( V,B)/T_K]$, the detailed behavior at $|\Delta \w| \sim D$
is not important. Similarly, it is sufficient to keep track only of
the real parts of the coupling constants on one Keldysh contour
denoted by $g^{\alpha\sigma\w_c; \alpha'\sigma'\w_c'}_{\gamma\w_f;
  \gamma'\w_f'}$ for an incoming electron in lead $\alpha=L,R$ with
energy $\w_c$ and spin $\sigma=\uparrow,\downarrow$ interacting with a
pseudo Fermion with spin $\gamma$ and frequency $\w_f$ describing the
local spin (see Fig.~\ref{dia}).  Primed quantities refer to  outgoing
particles. A calculation to higher order in $1/\ln[V/T_K]$ also has to
take into account the imaginary parts and the full Keldysh structure
of the vertices which can be neglected here.

From Fig.~\ref{dia} and Eq.~(\ref{logDer}) we obtain
\begin{multline}
\frac{\partial g^{\alpha\sigma,\w_c;
  \alpha'\sigma',\w_c'}_{\gamma,\w_f; \gamma',\w_f'}}{\partial \ln D}=
\frac{1}{8}\!
\sum_{\beta=L,R;\tau,\delta=\uparrow,\downarrow,\lambda=\pm}\biggl[\\
\  g^{\alpha\sigma,\w_c;
  \beta \tau, \mu_\beta+\lambda D}_{\gamma,\w_f;\delta,*}
\!  g^{ \beta \tau, \mu_\beta+\lambda D;
  \alpha'\sigma',\w_c'}_{\delta,*; \gamma',\w_f'}
        \,  \Theta_{\w_c+\w_f-\mu_{\beta}+ \delta B/2}  \\
-\ g^{\alpha\sigma,\w_c;
  \beta \tau, \mu_\beta+\lambda D}_{\delta, *; \gamma',\w_f'}
\! g^{ \beta \tau, \mu_\beta+\lambda D;
  \alpha'\sigma',\w_c'}_{\gamma,\w_f;\delta, *}
       \, \Theta_{\w_f'-\w_c+\mu_{\beta}+\delta B/2}  \biggr] \label{fullRG}
\end{multline}
where
\begin{equation}
\Theta_\w=\Theta(D-|\w|).
\end{equation}
 To simplify the notation, ``$*$'' represents in each term
the one frequency, which is fixed by energy conservation. The
initial conditions for Eqs.~(\ref{fullRG}) at the bare cutoff are
$g^{\alpha\sigma\w_c;
  \alpha'\sigma'\w_c'}_{\gamma\w_f; \gamma'\w_f'}=
\Nf J_{\alpha \alpha'} {\boldsymbol \tau}_{\sigma \sigma'}
{\boldsymbol \tau}_{\gamma \gamma'}$.

The renormalization group equations (\ref{fullRG}) are rather
complex and difficult to solve as each vertex depends on three
frequencies. Fortunately, one can drastically simplify them to
leading order in $1/\ln[\max( V,B)/T_K]$ using
 two approximations. First,  the energy of the spin is well defined
and the pseudo fermion spectral functions therefore strongly peaked
at $\omega=\pm B/2$, which allows setting $\omega_f$ ($\omega{'}_f$)
to $-\gamma\frac{B}{2}$ ($-\gamma{'}\frac{B}{2}$). Hence, one has to
keep track only of a {\em single} frequency, the energy of the
incoming electron. Second, the coupling functions appearing on the
right-hand side of (\ref{fullRG}) depend only logarithmically on the
frequency while the Heaviside step function is strongly frequency
dependent. This allows to neglect the frequency dependence of the
vertices by approximating $f(\Delta \w) \Theta(D-|\Delta \w|)\approx
f(0)\Theta(D-|\Delta \w|)$ on the right-hand side of (\ref{fullRG}).

It is also useful to introduce a more compact notation, which
separates spin-flip processes $g_\perp$ from non-flip vertices
$g_\|$ by defining
\begin{multline}
g^{\alpha\sigma,\w;
  \alpha'\sigma',\w- \frac{(\gamma-\gamma') B}{2}}_{\gamma,-\frac{\gamma B}{2}; \gamma',-\frac{\gamma' B}{2}}
=\tau_{\sigma \sigma'}^z \tau_{\gamma \gamma'}^z g_{z\sigma}^{\alpha \alpha'}(\w) \\
+(\tau_{\sigma \sigma'}^x \tau_{\gamma \gamma'}^x+\tau_{\sigma
\sigma'}^y \tau_{\gamma \gamma'}^y)g_{\perp \sigma}^{\alpha
\alpha'}(\w-\gamma \frac{B}{2}) \label{gs}
\end{multline}
The energy argument in the definition of $g_\perp$ has been chosen
to simplify the RG equations (see below). It turns out that the
property $g^{\uparrow \uparrow}_{\uparrow \uparrow}=-g^{\uparrow
\uparrow}_{\downarrow \downarrow}$ is preserved under RG, therefore
it is not necessary to keep track of a possible  $\gamma$ dependence
of $g_z$. Using these definitions and the approximations described
above one obtains
\begin{multline}
\frac{\partial g^{\alpha\alpha'}_{\perp \sigma}(\w)}{\partial \ln
D}= -\frac{1}{2}\!
\sum_{\beta=L,R}\biggl[\\
g^{\alpha\beta}_{\perp \sigma}(-\frac{\sigma B}{2}-\frac{\beta
V}{2})g^{\beta\alpha'}_{z,-{\sigma}}(-\frac{\beta V}{2})
 \Theta_{\w+\frac{\sigma B}{2}+\frac{\beta V}{2}}\\
+g^{\alpha\beta}_{z \sigma}(-\frac{\beta
V}{2})g^{\beta\alpha'}_{\perp {\sigma}}(\frac{\sigma
B}{2}-\frac{\beta V}{2})
 \Theta_{\w-\frac{\sigma B}{2}+\frac{\beta V}{2}}\biggr] \label{RG1}
\end{multline}
and for the longitudinal component
\begin{multline}
\frac{\partial g^{\alpha\alpha'}_{z \sigma}(\w)}{\partial \ln D}=
-\!
\sum_{\beta=L,R}\biggl[\\
g^{\alpha\beta}_{\perp \sigma}(-\frac{\sigma B}{2}-\frac{\beta
V}{2})g^{\beta\alpha'}_{\perp,-{\sigma}}( -\frac{\sigma
B}{2}-\frac{\beta V}{2})
 \Theta_{\w+\sigma B+\frac{\beta V}{2}} \biggr] \label{RG2}
\end{multline}
 The initial values of the RG equations are
$g^{\alpha\alpha'}_{z \sigma}(\w)=g^{\alpha\alpha'}_{\perp
\sigma}(\w)=g^{\alpha \alpha'}$.

The equations (\ref{RG1}) and (\ref{RG2}) together with the results
for the dephasing rates and other physical quantities given below
are the main result of this paper. In Ref.~\cite{Rosch03a}, these
equations have been presented for the special case $g^{\alpha
\alpha'}=const.$ independent of the lead index, where they simplify
drastically (see below).

It is useful to discuss the symmetries of Eq.~(\ref{RG1}) and
(\ref{RG2}). Due to the hermiticity of the underlying problem one
finds
\begin{eqnarray} \label{sym1}
g^{\alpha\alpha'}_{\perp \sigma}(\w)&=&g^{\alpha'\alpha}_{\perp,-\sigma}(\w) \label{gperpS}\\
g^{\alpha\alpha'}_{z \sigma}(\w)&=&g^{\alpha'\alpha}_{z \sigma}(\w).
\end{eqnarray}
 The coupling matrices simplify drastically in certain limits. For
example, if the Kondo model is derived from an underlying Anderson
model, then only a single channel couples to the dot and the matrix
of bare couplings $g^{\alpha \alpha'}$ has only a single
non-vanishing eigenvalue. This property is conserved under RG as the
RG equations have the matrix structure $\partial \hat{g}=\hat{g}
\hat{M} \hat{g}$ and therefore one can express the matrices in LR
space in terms of
   a single function $g_{\perp/z}(\omega)$
\begin{eqnarray}\label{symch}
g^{\alpha \alpha'}_{\perp/z}(\w)= 2 g_{\perp/z}(\w)
\left(\begin{array}{c} \cos \phi\\\sin \phi
\end{array}\right)  (\cos \phi,\sin \phi) \label{gSi}
\end{eqnarray}
Due to (\ref{gperpS}), $g_\perp$ does not depend on $\sigma$ in this
case, $g_{\perp \uparrow}(\w)=g_{\perp \downarrow}(\w)=g_\perp(\w)$.

In equilibrium, $V=0$, it is well known that within one-loop RG, the
two eigenvalues of the coupling constant matrix $\hat{g}=g^{\alpha
\alpha'}$ do not mix. This is therefore also a property of the RG
equations (\ref{RG1}) and (\ref{RG2}), which have in this case the
matrix structure $\partial \hat{g} \propto \hat{g} \hat{g}$. For
each of the eigenvalues of $\hat{g}$ one obtains for $V=0$
\begin{eqnarray}
\frac{\partial g_{z \sigma}(\w)}{\partial \ln D}&=& -2
(g_\perp(B/2))^2 \Theta_{\w+\sigma B}
\\
\frac{\partial g_{\perp}(\w)}{\partial \ln D}&=&
- g_\perp(B/2) g_z(0)\left[\Theta_{\w+B/2} + \Theta_{\w-B/2}\right]\nonumber \\
\end{eqnarray}
In Ref.~\cite{spectral} we used these equations to calculate the
spectral functions at large frequencies and magnetic fields. We
checked that these equations indeed reproduce the perturbation
theory for the T-matrix not only to  order $g^2 \ln[..]$ but also to
order $g^3 \ln^2[..]$ as has to be expected from a theory which
resums leading logarithmic corrections. Furthermore, we
compared\cite{spectral} our results to spectral functions calculated
with numerical renormalization group and found that the relative
errors are -- again as expected -- of order $1/\ln[\max(\w,B)/T_K]$.

For vanishing magnetic field $B$ and spin-rotational invariant bare
couplings $J$, the renormalized coupling constants obviously do not
depend on spin and the RG equations simplify to
\begin{multline}\label{RGV}
\frac{\partial g^{\alpha \alpha'}(\w)}{\partial \ln D}=
-\sum_{\beta=L,R} g^{\alpha \beta}\!(-\frac{\beta V}{2}) \, g^{
\beta \alpha'}\!(-\frac{\beta V}{2})  \Theta_{\w+\beta V/2}.
\end{multline}

If the dot couples symmetrically to the left and right lead, one has
\begin{eqnarray}\label{symsym}
g_{\perp/z, \sigma}^{LL}(\w)&=&g_{\perp/z,- \sigma}^{RR}(-\w).
\end{eqnarray}
If, furthermore, only a single channel couples symmetrically to the
dot, one can use (\ref{gSi}) to parameterize the renormalized
vertices by just two functions $g_\perp(\w)=g_\perp(-\w)$ and $g_{z
\uparrow}(\w)=g_{z \downarrow}(-\w)$ which are calculated from the
RG equations
\begin{eqnarray}
\frac{\partial g_{z\sigma}(\w)}{\partial \ln D} &=&
-\sum_{\beta=-1,1} g_\perp(\frac{B+\beta V}{2})^2 \Theta_{\w+\sigma
(B+ \beta \frac{V}{2})}
 \label{jRG} \\
\frac{\partial g_{\perp}(\w)}{\partial \ln D} &=& -\!\!\!
\sum_{\beta,\sigma=-1,1} \!\!\! \frac{g_\perp(\frac{\sigma B+\beta
V}{2}) g_{z\sigma}(\frac{\beta V}{2})}{2} \Theta_{\w+\frac{\sigma
B+\beta V}{2}} \nonumber
\end{eqnarray}
This limit has been discussed before in Ref.~\cite{Rosch03a}.

How are physical quantities calculated from the renormalized
coupling constants? To leading order in our small parameter
$1/\ln[\max(V,B,\w)/T_K]$ it turns out that it is sufficient to
replace in the 2nd order perturbation theory expressions (or,
equivalently, in the golden rule expressions) the bare coupling
constants by the appropriate renormalized couplings in the limit
$D\to 0$. In each case we have checked this by comparing with
perturbation theory to order $g^3$, where the first logarithmic
correction arises\cite{paaske1}. Experimentally the most relevant
quantity is the current
\begin{multline}\label{current}
I=\frac{2 \pi e}{16 \hbar} \sum_{\gamma \sigma} \int d\w \, f^L_\w (1-f^R_\w) n_\gamma [g_{z\sigma}^{LR}(\w)]^2\\
+ 4  f^L_\w (1-f^R_{\w-\gamma B}) n_\gamma
[g_{\perp,-\gamma}^{LR}(\w-\frac{\gamma B}{2})]^2 - (L
\leftrightarrow R)
\end{multline}
where $f^{L/R}_\w=f(\w \mp V/2)$ are the Fermi functions in the left
and right leads. The spin occupations $n_{\uparrow/\downarrow}$ are
calculated from the rate equation
\begin{equation}\label{ratee}
\Gamma_{\downarrow \to \uparrow}=\Gamma_{\uparrow \to \downarrow}
\end{equation}
 [see Eq.~(\ref{boltz})]
with $n_{\uparrow}+n_{\downarrow}=1$ and
\begin{eqnarray}\label{rate1}
\Gamma_{\uparrow \to \downarrow}\! &=& \! \frac{2 \pi}{4 \hbar}
n_\uparrow \sum_{\alpha \alpha'} \int \! d\w \, f^\alpha_\w
(1-f^{\alpha'}_{\w-B}) [g_{\perp \downarrow}^{\alpha
\alpha'}(\w-\frac{B}{2})]^2
\nonumber\\
\Gamma_{\downarrow \to \uparrow}\! &=& \! \frac{2 \pi}{4 \hbar} n_\downarrow \sum_{\alpha \alpha'} \int \! d\w \, f^\alpha_\w (1-f^{\alpha'}_{\w+B}) [g_{\perp \uparrow}^{\alpha \alpha'}(\w+\frac{B}{2})]^2. \nonumber \\
\end{eqnarray}
A quantity of interest for spintronics applications is  the
spin-current\cite{gabriel} $I_\text{spin}=-\frac{1}{2}
(\dot{S}_{Lz}-\dot{S}_{Rz})$, where $S_{L/R,z}$ is the $z$ component
of the total spin of the conduction electrons in the left/right
lead.
\begin{multline}\label{spincurrent}
I_\text{spin}=\frac{ \pi }{16 \hbar} \sum_{\gamma \sigma \alpha \alpha'} \int d\w \, \sigma \frac{\alpha'-\alpha}{2} f^\alpha_\w (1-f^{\alpha'}_\w) n_\gamma [g_{z\sigma}^{\alpha \alpha'}(\w)]^2\\
+4 n_\gamma  f^\alpha_\w (1-f^{\alpha'}_{\w-\gamma B})
\frac{\alpha+\alpha'}{2} [g_{\perp,-\gamma}^{\alpha
\alpha'}(\w-\frac{\gamma B}{2})]^2
\end{multline}
Spin transport will be discussed in detail elsewhere\cite{spinpub}.

As we argued in the introduction, the RG flow will finally be cut off
and controlled by the dephasing of coherent spin-flips. Therefore
one also has to calculate the dephasing rate $\Gamma=1/T_2$ by
replacing the bare couplings in the perturbative
formula\cite{paaske2} by the renormalized couplings
\begin{multline}\label{gamma}
\Gamma=\frac{1}{T_2}=\frac{\pi}{8 \hbar} \sum_{\gamma \sigma \alpha
\alpha'} \int d\w \,
f^\alpha_\w (1-f^{\alpha'}_{\w})  [g_{z\sigma}^{\alpha \alpha'}(\w)]^2\\
+  f^\alpha_\w (1-f^{\alpha'}_{\w-\gamma B})
[g_{\perp,-\gamma}^{\alpha \alpha'}(\w-\frac{\gamma B}{2})]^2
\end{multline}
This formula takes into account a partial cancelation of self-energy
and vertex corrections. In a careful analysis of perturbation
theory, it was shown in Ref.~\cite{paaske2}, how decoherence rates
cut off logarithmic singularities and therefore stop the
renormalization group flow.  It turns out~\cite{paaske2} that
 different rates (combinations of $1/T_1$ and $1/T_2$ in the language of NMR) enter the various logarithms.
For the situation discussed in this paper, however, the various
relaxation rates differ only by factors of order 1. As the
relaxation rates appear only (see below) in arguments of logarithms,
$\ln \Gamma/T_K$, it is not necessary to keep track of these
prefactors to leading order in $1/\ln[ \max(V,B,\w)/T_K]$. Hence, we
can approximate all relaxation rates by $\Gamma$. To leading order,
the effect that the self-energy and vertex corrections cut off all
logarithmic contributions at the scale $\Gamma$, can be taken into
account by replacing
\begin{equation}\label{thetaGamma}
\Theta_\omega=\Theta (D-\sqrt{\omega^2 + \Gamma^2})
\end{equation}
in (\ref{RG1},\ref{RG2}). A further effect of the relaxation rate is
that spectral functions and spin-spin correlation functions are
broadened on the scale $\Gamma$. This can effectively be described
by replacing $f_\w$ appearing in (\ref{current}--\ref{gamma}) by
Fermi functions smeared out on the scale $\Gamma$ (note, however,
that the distribution functions of electrons in the leads are not
renormalized). Details of how this broadening is justified  and
implemented will be published elsewhere\cite{preparation}.

A full solution of the RG equation can now be obtained in the
following way (an example is given below): First, the RG equations
(\ref{RG1},\ref{RG2}) are solved for $\Gamma=0$. This turns out to
be rather easy as on the right-hand side of the equations only
couplings at a finite set of fixed frequencies enter. One therefore
first solves for those frequencies and constructs the full energy
dependence in the limit $D\to 0$ in a second step. For some
frequencies one will find diverging couplings as the effect of
$\Gamma$ has not yet been taken into account. With these frequency
dependent couplings, one calculates $\Gamma$ from (\ref{gamma})
which is finite as the integrals over the weakly divergent coupling
functions do not diverge. With this $\Gamma$ one recalculates the RG
equations using (\ref{thetaGamma}). In practical numerical
implementations, we iterate this procedure until convergence is
reached, but formally the effects of self-consistency are subleading
and the first iteration described above is sufficient within the
precision of our approach. Finally, physical quantities like the
magnetization [using (\ref{rate1})], the current (\ref{current}) or
the spectral function \cite{spectral} are calculated from the
renormalized couplings. Note that the approximation, that $\Gamma$
and other physical quantities like the occupations can be calculated
in an independent second step after solving the RG equations, is
only valid to leading order in $1/\ln[\max(V,B)/T_K]$. To higher
order, one has to solve simultaneously the RG equations for
(frequency dependent) self-energies and vertices.

As an example, we consider a simple situation by calculating the
properties of  a symmetric dot with $g^{LL}=g^{RR}=g_d$ for
vanishing magnetic field, $B=0$, and large voltages, $V \gg T_K$.
From (\ref{sym1}) and (\ref{symsym}) one finds
$g^{LL}(\w)=g^{RR}(-\w)=g^d(\w)$ and
$g^{LR}(\w)=g^{RL}(\w)=g^{LR}(-\w)$. To solve the RG equations
(\ref{RGV}), we first observe that on their right hand side only the
couplings $g_d(V/2)$ and $g^{LR}(V/2)$ enter. Therefore we first
determine the RG equation for these two constants
\begin{eqnarray}
 \nonumber
  \partial g_d(V/2) &=& -[g^{LR}(V/2)^2 \Theta_V+g_d(V/2)^2 \Theta_\Gamma] \\
 \partial g^{LR}(V/2) &=& -g^{LR}(V/2) g_d(V/2)( \Theta_V+\Theta_\Gamma)
 \end{eqnarray}
with $\Theta_\Gamma=\Theta[D-\Gamma]$ and $\Theta_V\approx
\Theta[D-V]$ as $\Gamma \ll V$. Note, that the RG flow is {\em not}
completely cutoff by $V$ but continues for $D>\Gamma$. The equations
are most conveniently solved in the scaling limit, where the Kondo
temperature is held fixed assuming a large initial cutoff $D \to
\infty$ and a small bare exchange coupling $J\to 0$. For $D>V$, the
RG equations are identical to the RG equations for the two-channel
channel-anisotropic Kondo model in equilibrium, the channels are the
even (L+R) and odd (L-R) linear combinations of electrons. The RG
flow is characterized by two RG invariants, the Kondo temperature
$T_K= D e^{-1/(g_d+g^{LR})}$ and $\alpha_0=\frac{g_d^2-(g^{LR})^2}{
2 g^{LR}}$. The RG equations are easily solved and one obtains
$g_d(V/2)=\frac{1+2 \alpha_0 \ln[D/T_K]}{2 \ln[D/T_K] (1+\alpha_0
\ln [D/T_K])}$ for $D>V$ and $g_d(V/2)=1/\ln[D/T^*]$ with $T^*=T_K
(T_K/V)^{1/(1+2 \alpha_0 \ln[V/T_K])}$ for $\Gamma<D<V$. The
coupling $g_{LR}(V/2)$ is given by $g_{LR}(V/2)=g_d(V/2)/(1+2
\alpha_0 \ln[\max(D,V)/T_K])$. Plugging these solutions into the
right-hand side of (\ref{RGV}), one can easily integrate the
equations analytically for an arbitrary frequency. As the formulas
are rather lengthy,  we show only the result for $\alpha_0=0$ when
only a single channel couples to the dot and
$g_d(\w)=g^{LR}(\w)=g(\w)$ according to Eq.~(\ref{symch}). For $D\to
0$ one finds
\begin{multline}\label{gdw}
    g(\w)\approx \sum_\beta \Theta[|\w-\beta V/2|-V] \frac{1}{4 \ln[|\w-\beta V/2|/T_K]}\\+
    \Theta[V-|\w-\beta V/2|] \left(\frac{1}{\ln[V \max(|\w-\beta V/2|,\Gamma)/T_K^2]}\right.\\\left.-\frac{1}{4 \ln[V/T_K]}\right)
\end{multline}
where $\Gamma\approx V \frac{\pi}{16 \ln[V/T_K]^2}$ is calculated
from (\ref{gamma}). In Fig.~\ref{dia} a plot of $g(\w)$ is shown for
$V/T_K=100$, where $\Gamma \approx 1.6 T_K$. For $V \gg T_K$, the
relaxation rate is larger than $T_K$ and therefore one stays in the
weak coupling regime, $g(\w) \ll 1$ for $\alpha_0=0$. As described
in detail in Ref.~\cite{Rosch01}, for $\alpha_0>1$ the perturbative
RG breaks down for large voltages and one enters a strong coupling
regime. (In that analysis\cite{Rosch01} we did not keep track of the
frequency dependent couplings, nevertheless all results remain
valid; $g_{LR}$ in Ref.~\cite{Rosch01} has to be identified with
$g_{LR}(\w=0)$).

Using (\ref{gdw}) one can easily calculate physical quantities like
the current. As the current is obtained from an integral over
$g(\w)$ it is not sensitive to details of $g(\w)$ and one finds that
the differential conductance is given by $G\approx \frac{3 \pi e
V}{16 \hbar \ln[V/T_K]^{2}}$, a result obtained before by Kaminski {\it
et al.}\cite{Kaminski99}. A quantity like the susceptibility,
however, is sensitive to the peaks in $g(\w)$ and one finds
(restoring the $\alpha_0$ dependence) by solving the rate equation
(\ref{ratee})
\begin{eqnarray}
\chi(V)=\frac{2}{V} \frac{1+2 \alpha_0 \ln \frac{V}{T_K} (1+
\alpha_0 \ln \frac{V}{T_K})} {\left( 1-
 \frac{(1 + 2 \alpha_0 \ln \frac{V}{T_K}) \ln\left[ \ln \frac{V}{T_K}
(1+\alpha_0  \ln \frac{V}{T_K})\right]   }{(1 +  \alpha_0 \ln
\frac{V}{T_K}) \ln  \frac{V}{T_K}} \right)^2}
\end{eqnarray}
While in the presence of a finite bias voltage, large logarithmic
corrections arise [related to the peaks in $g(\w)$], one obtains in
equilibrium to the same order of approximation,  just $\chi=1/(2 T)$.
Expanding this result in the bare couplings, one obtains
Eq.~(\ref{eq:chi}).

\begin{figure}
\includegraphics[width=0.9  \linewidth]{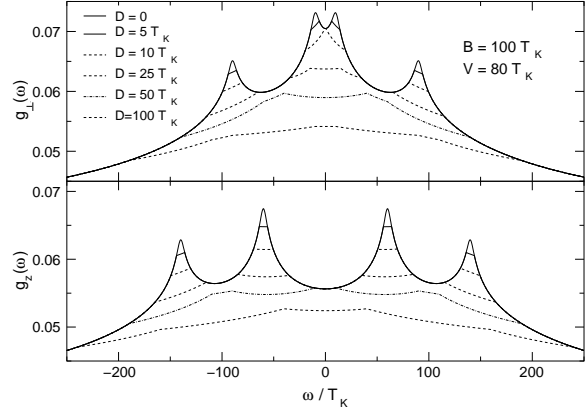}
\caption{\label{figJ} Renormalized coupling constants $g_\perp(\w)$
(upper panel) and $g_z(\w)$ (lower panel) for $B=100 T_K$ and $V=80
T_K$ and for different values of the running cut-off $D$.  A
symmetric dot with bare couplings $g^{\alpha \alpha'}=g$ is assumed.
Peaks signal that resonant scattering processes take place at those
frequencies. The peaks are cutoff by the relaxation rate $\Gamma$.}
\end{figure}

In Fig.~\ref{figJ}, we show the renormalized coupling constants for
various values of the running cutoff $D$ in the presence of both a
bias voltage $V$ and a magnetic field $B$. While asymptotically the
RG equations can be solved analytically, the various crossover
regimes have to be calculated from a numerical solution following
the approach described above. When the cutoff $D$ is lowered, more
and more scattering processes freeze out and only resonant
contributions with $\w\approx \pm V/2 \pm B/2$ or $\w \approx \pm
V/2\pm B$ survive and lead to pronounced peaks in the renormalized
coupling constants.

\begin{figure}[t]
\begin{center}
\includegraphics[width=0.7 \linewidth]{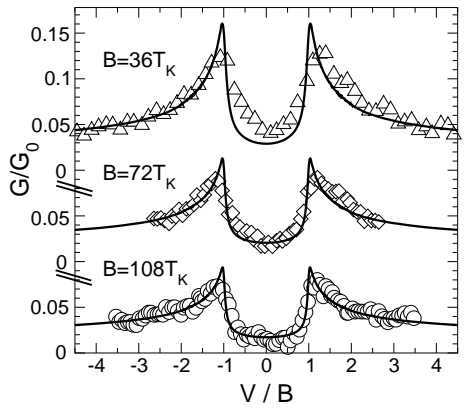}
\end{center}
 \vspace*{-.3cm}
\caption{\label{figG} Conductance measurements
  of Ref.\ \cite{Ralph94} (symbols) on metallic point contacts
  in magnetic fields $0.85,1.7,2.55\,$T ($B=36,72,104\, T_K$,  with
  $T_K\approx30$mK \cite{Ralph94}).  Assuming that the corresponding
  point contact is described by a single-channel ($J_{LR}=\sqrt{J_{LL}
    J_{RR}}$) Kondo model, $\frac{J_{RR}}{J_{LL}}\approx
  4.2$ is determined from $G(V\!\!=\!\!0,B\!\!=\!\!0,T\!\!=\!\!50
  \text{mK})=\frac{4 J_{LL} J_{RR}}{(J_{LL}+J_{RR})^2}
  G_{\text{sym}}(T/T_K)$, where $G_{\text{sym}}$ is known exactly from
  NRG calculations \cite{Costi94}.  This fixes {\em all} parameters for
  our RG calculation (solid lines, $T=0$).
  As the ($B$-dependent) background  is not known experimentally,
  we subtract $\Delta G=G_B-5.2\cdot 10^{-5} G_0 \frac{V}{T_K}$, where
  $G_B$ is fitted to our results at large $V$.
  The experimental temperature
  $T=50$mK leads to an extra small broadening at $V=B$. Note that the agreement between theory and experiment is best for large voltages and magnetic fields as expected.}
\end{figure}

Experimentally, the best accessible quantity is the differential
conductance $G(V,B)$ as a function of bias voltage and magnetic
field. Our calculations can be applied to all systems described by
the simple Kondo model (\ref{H}), e.g. quantum dots with an odd
number of electrons in the Coulomb blockade regime, provided that
either voltages or magnetic fields are large compared to the Kondo
temperature, $V,B \gg T_K$, and provided that other excitations  of
the dot have a much higher energy. The latter requirement is often
not fulfilled in quantum dots or molecules. While a generalization
of the above described methods to more complex dots and molecules
with more levels is straightforward, this will introduce more
fitting parameters. If the Kondo model (\ref{H}) is valid, just two
parameters are needed: the Kondo temperature $T_K$ and the L/R
asymmetry of the coupling $J_{LL}/J_{RR}$ (assuming that only a single
channel couples to the Kondo spin as it is the case in most single
dot experiments). Both parameters can be determined by comparing the
$T$ dependence of the linear-response conductance $G(T,V\to 0,B=0)$
to exact theoretical results determined from  numerical
renormalization group calculations\cite{Costi94}. This allows for a
parameter-free theoretical prediction of $G(V/T_K,B/T_K)$ for large
$V$ and $B$. In Fig.~\ref{figG} we compare\cite{Rosch03a}  our
theoretical results [determined from a numerical solution of
Eqs.~(\ref{RG1},\ref{RG2},\ref{current},\ref{rate1},\ref{gamma})
using (\ref{symch})] to experiments by Ralph and
Buhrman\cite{Ralph94}, where ratios of $V/T_K$ and $B/T_K$ larger
than 100 have been probed. The peaks in the differential conductance
$G(V/B,B/T_K)$ at $V\sim B$ and their characteristic shape arises
due to the interplay of several effects. For $V \ll B$ and $B \gg
T_K$, the spin is strongly polarized, the Kondo effect is suppressed
and transport is dominated by processes where the spin does not
flip. For $V\gtrsim B$ a new transport channel is opened, as the
voltage provides a sufficient amount of energy to flip the spin.
Simultaneously, the magnetization on the dot changes and resonant
spin-flip scattering from one Fermi surface to the other strongly
renormalizes the effective couplings. The smooth drop at large
voltages arises, because the Kondo effect is further suppressed by
the voltage. The theory fits the experiment surprisingly well,
especially when one takes into account that a relative error of
order $1/\ln[\max(V,B)/T_K]$ has to be  expected from our
calculation. Either the next-order correction is accidentally small
or some error in the determination of $T_K$ (we took the value
quoted in the experimental paper\cite{Ralph94}) accidentally
improves the fit. It is, however, worthwhile to stress, that
deviations of theory and experiments are strongest in the regime of
small voltages and magnetic fields as expected from an approach
based on an expansion in $1/\ln[\max(V,B)/T_K]$.

\section{Conclusions and Outlook}

In the last 40 years the Kondo effect has motivated many new
theoretical developments. It can be expected that the Kondo model
will also play an important role in developing techniques to
understand and describe quantum systems out of equilibrium in
regimes where perturbative methods are not available. For example,
it would be extremely useful to establish novel numerical methods
which are able to calculate e.g. the current-voltage characteristic
of a quantum-dot in the Kondo regime. Besides the obvious practical
use of methods to predict and explain transport properties of
single-electron transistors, molecules and similar systems, we
believe that it will be also a conceptual challenge to develop the
language and tools to describe and classify strongly interacting
quantum systems in steady-state non-equilibrium.

In this paper, we have  discussed a simple approach to generalize
perturbative renormalization group to a situation, where a finite
bias voltage is applied to a quantum dot in the Kondo regime. We
derived the one-loop RG equations for a spin coupled to the
electrons in two leads by an arbitrary matrix of exchange couplings.
This allows a controlled calculation of spectral functions,
differential conductances, susceptibilities and other properties of
the dot for large bias voltages and/or large magnetic fields and
frequencies. The expansion is controlled by the small parameter
$1/\ln[\max(V,B,\w)/T_K]$. For the future, it will be interesting to
generalize this approach to more complex situations and to
investigate  systems where the perturbative renormalization group
breaks down at large bias voltages\cite{Rosch01}.

\section*{Acknowledgment}
We thank T. Costi, S. Franceschi M. Garst, L. Glazman, H. Schoeller
and M. Vojta for useful discussions. Part of this work was done at
the Aspen Center of Physics.

%\appendix
%\section{sample}


\begin{thebibliography}{99} %% The number "99" means that this list has more than nine items.
%\bibitem{jjap} S. Nakamura: Jpn. J. Appl. Phys. \textbf{30} (1991) L1705.
%\bibitem{jpsj} J. Akimitsu, H. Ichikawa, N. Eguchi, T. Miyano, M. Nishi and K. Kakurai: J. Phys. Soc. Jpn. \textbf{70} (2001) 3475.
%\bibitem{bcsj} Y. Mizutani and T. Kiatgawa: Bull. Chem. Soc. Jpn. \textbf{75}  (2002) 623.
%\bibitem{apl} S. F. Chichibu, A. Setoguchi, A. Uedono, K. Yoshimura and M. Sumiya: Appl. Phys. Lett. \textbf{78} (2001) 28.
%\bibitem{jap}  Y. Ikeda, K. Suzuki, H. Fukumoto, J. P. Verboncoeur, P. J. Christenson,
% C. K. Birdsall, M. Shibata and M. Ishigaki: J. Appl. Phys. \textbf{88} (2000) 6216.
%\bibitem{prb} N. Harima, J. Matsuno, A. Fujimori, Y. Onose, Y. Taguchi and Y. Tokura: Phys. Rev. B \textbf{64} (2001) 220507(R).
%\bibitem{prl} K. Akama, T. Hattori and K. Katsuura: Phys. Rev. Lett. \textbf{88} (2002) 201601.
%\bibitem{jcp} A. Nakayama and K. Yamashita: J. Chem. Phys. \textbf{114} 780.
%\bibitem{jcg} I. Ohkubo, Y. Matsumoto, K. Ueno, T. Chikyow, M. Kawasaki and H. Koinuma: J. Cryst. Growth \textbf{247} (2001) 105.
%\bibitem{ed} Y. Negoro, N. Miyamoto, T. Kimoto and H. Matsunami: IEEE Electron Devices \textbf{49} (2002) 1505.


\bibitem{Kondo64}J. Kondo:
Prog. Theor. Phys. {\bf 32}, 37 (1964).

\bibitem{Hewson93}A. C. Hewson: {\it The Kondo Problem to Heavy Fermions}
(Cambridge University Press, 1993).


\bibitem{Appelbaum72}J. Appelbaum and L. Y. Shen
Phys. Rev. B {\bf 5}, 544 (1972).

\bibitem{Wallis74}R. H. Wallis and A. F. G. Wyatt:
J. Phys. C: Solid State Phys. {\bf 7}, 1293 (1974).

\bibitem{Wolf70}E. L. Wolf and D. L. Losee:
Phys. Rev. B {\bf 2}, 3660 (1970).

%\bibitem{Wolf75}E. L. Wolf: in {\it Solid State Physics}, eds. H. Ehrenreich,
%F. Seitz and D. Turnbull (Academic, New York, 1975), vol. 30, p.1.5

\bibitem{Shen68}L. Y. L. Shen and J. M. Rowell,
Phys. Rev. {\bf 165}, 566 (1968).

\bibitem{Nielsen70}P. Nielsen:
Phys. Rev. B {\bf 2}, 3819 (1970).

\bibitem{Bermon78}S. Bermon, D. E. Paraskevopoulos and P. M. Tedrow:
Phys. Rev. B {\bf 17}, 210 (1978).

\bibitem{Appelbaum66}J. Appelbaum:
Phys. Rev. Lett. {\bf 17}, 91 (1966).

\bibitem{Appelbaum67a}J. Appelbaum:
Phys. Rev. {\bf 154}, 633 (1967).


\bibitem{Anderson66}P. W. Anderson:
Phys. Rev. Lett. {\bf 17}, 95 (1966).


%\bibitem{Losee69}D. L. Losee and E. L. Wolf:
%Phys. Rev. Lett. {\bf 23}, 1457 (1969).

%\bibitem{Ivezic80}T. Ivezi\'{c}:
%J. Magn. Magn. Mater. {\bf 15-18}, 933 (1980).


\bibitem{Solyom68}J. S\'{o}lyom and A. Zawadowski:
Phys. Kondens. Materie {\bf 7}, 325 (1968); Phys. Kondens. Materie
{\bf 7}, 342 (1968).

\bibitem{Appelbaum70}J. Appelbaum and W. F. Brinkman
Phys. Rev. B {\bf 2}, 907 (1970).

\bibitem{Ivezic75}T. Ivezi\'{c}:
J. Phys. C: Solid State Phys., {\bf 8}, 3371 (1975).

\bibitem{Wolf85}E. L. Wolf: {\it Principles of Electron Tunneling
    Spectroscopy}, (Oxford University Press, Oxford, 1985).



\bibitem{Goldhaber98}D. Goldhaber-Gordon: Hadas Shtrikman: D. Mahalu:
David Abusch-Magder: U. Meirav and M. A. Kastner: Nature {\bf 391},
156 (1998).

\bibitem{Cronenwett98} S. M. Cronenwett, T. H. Oosterkamp and L. P. Kouwenhoven:
Science {\bf 281}, 540 (1998).

\bibitem{Schmid98} J. Schmid, J. Weis, K. Eberl and K. von Klitzing:
Physica {\bf B258}, 182 (1998).

\bibitem{Nygaard00}J. Nyg{\aa}rd, D. H. Cobden and P. E. Lindelof:
Nature {\bf 408}, 342 (2000).

\bibitem{vanderWiel00}W. G. van der Wiel, S. De Franceschi, T. Fujisawa,
J. M. Elzerman, S. Tarucha, and L. P. Kouwenhoven: Science {\bf
289}, 2105 (2000).

\bibitem{Goldhaber98b}D. Goldhaber-Gordon, J. G\"{o}res, M. A. Kastner,
Hadas Shtrikman, D. Mahalu and U. Meirav: Phys. Rev. Lett. {\bf 81},
5225 (1998).


\bibitem{Glazman88}L. Glazman and M. Raikh:
JETP Letters {\bf 47}, 452 (1988).

\bibitem{Ng88} T. Ng and P.A. Lee:
Phys. Rev. Lett. {\bf 61}, 1768 (1988).


\bibitem{Ralph94} D. C. Ralph and R. A. Buhrman, Phys. Rev. Lett. {\bf 72}, 3401 (1994).

% Out-of-Equilibrium Kondo Effect in a Mesoscopic Device
\bibitem{noneqDistr}
S. De Franceschi, R. Hanson, W. G. van der Wiel, J. M. Elzerman, J.
J. Wijpkema, T. Fujisawa, S. Tarucha, and L. P. Kouwenhoven: Phys.
Rev. Lett. 89, 156801 (2002).


\bibitem{Goldin98}Y. Goldin and Y. Avishai:
Phys. Rev. Lett. {\bf 81}, 5394 (1998); Phys. Rev. B {\bf 61}, 16750
(2000).

\bibitem{Appelbaum67b}J. Appelbaum, J. C. Phillips and G. Tzouras:
Phys. Rev. {\bf 160}, 554 (1967).

\bibitem{Kaminski99}A. Kaminski, Yu. V. Nazarov and L. I. Glazman:
Phys. Rev. Lett {\bf 83}, 384 (1999); Phys. Rev. B {\bf 62}, 8154
(2000).

\bibitem{hettler95}
M. H. Hettler and H. Schoeller: Phys. Rev. Lett. {\bf 74}, 4907
(1995).

%ac Response in the Nonequilibrium Anderson Impurity Model
\bibitem{ng96} T-K. Ng: Phys. Rev. Lett. {\bf 76}, 487 (1996).

%Kondo Effect in ac Transport through Quantum Dots
\bibitem{lopez}
R. Lopez, R. Aguado, G. Platero, and C. Tejedor: Phys. Rev. Lett.
{\bf 81}, 4688 (1998).



%Photon-Induced Kondo Satellites in a Single-Electron Transistor
\bibitem{kogan} A. Kogan, S. Amasha, and M. A. Kastner: Science {\bf
304} 1293 (2004).

%%%%%%%% spins in a wire %%%%%%%%%%%%%

\bibitem{pothier.97} H. Pothier,
S. Gu\'eron, Norman. O. Birge, D. Esteve, and M. H. Devoret, Phys.
Rev. Lett. {\bf 79}, 3490 (1997).
%%Z. Phys. B {\bf 104}, 178 (1997).

\bibitem{anthore.03}
A. Anthore, F. Pierre, H. Pothier, and D. Esteve, Phys. Rev. Lett.
{\bf 90}, 076806 (2003).

\bibitem{schopfer.03}
F. Schopf, C. B\"auerle, W. Rabaud, and L. Saminadayar, Phys. Rev.
Lett. {\bf 90}, 056801 (2003).


\bibitem{glazman.01}
A. Kaminski and L.I. Glazman, Phys.~Rev.~Lett. {\bf 86}, 2400
(2001).

\bibitem{goeppert.01}
G. G\"oppert, H. Grabert, Phys. Rev. B {\bf 64}, 033301 (2001).

\bibitem{kroha.02}
J. Kroha and A. Zawadowski, Phys. Rev. Lett. {\bf 88}, 176803
(2002).

%%%%%%%%
\bibitem{Sivan96}N. Sivan and N. S. Wingreen:
Phys. Rev. B {\bf 54}, 11622 (1996).

 \bibitem{hersh91} S. Hershfield, J. D. Davies, and J. W. Wilkins: Phys. Rev. Lett. {\bf 67}, 3720 (1991).

\bibitem{Fujii} T. Fujii and K. Ueda:
%Perturbative approach to the nonequilibrium Kondo effect in a quantum dot
Phys. Rev. B {\bf 68} (2003) 531.

\bibitem{Oguri} A. Oguri:
%Out-of-equilibrium Anderson model at high and low bias voltages
  J. Phys. Soc.  J. {\bf 71} (2002) 2969.

\bibitem{Rosch03a}A. Rosch, J. Paaske, J. Kroha and P. W\"olfle:
Phys. Rev. Lett. {\bf 90}, 076804 (2003)

\bibitem{paaske1}
%Nonequilibrium Transport through a Kondo Dot in a Magnetic Field:
%Perturbation Theory
J. Paaske, A. Rosch, P. W\"olfle: Phys. Rev. B {\bf 69}, 155330
(2004).

 \bibitem{paaske2}
%Nonequilibrium Transport through a Kondo Dot: Decoherence Effects
J. Paaske, A. Rosch, J. Kroha, and P. W\"olfle: Phys. 
Rev. B. {\bf 70}, 155301 (2004).

\bibitem{Parcollet02}O. Parcollet and C. Hooley:
Phys. Rev. B {\bf 66}, 085315 (2002).


\bibitem{Meir93}Y. Meir, N. S. Wingreen and P. A. Lee:
Phys. Rev. Lett. {\bf 70}, 2601 (1993).

\bibitem{Wingreen94}N.S. Wingreen and Y. Meir:
Phys. Rev. B {\bf 49}, 11040 (1994).

\bibitem{Hettler94}M. H. Hettler, J. Kroha and S. Hershfield:
Phys. Rev. Lett. {\bf 73}, 1967 (1994); Phys. Rev. B {\bf 58}, 5649
(1998).

\bibitem{Konig96}J. K\"{o}nig, J. Schmid, H. Schoeller and G. Sch\"{o}n,
Phys. Rev. B {\bf 54}, 16820 (1996). 

\bibitem{Norlander99}P. Nordlander, M. Pustilnik, Y. Meir, N. S. Wingreen and
D. C. Langreth: Phys. Rev. Lett. {\bf 83}, 808 (1999);

\bibitem{Plihal00}M. Plihal, D. C. Langreth, P. Nordlander:
Phys. Rev. B {\bf 61}, R13341 (2000);


 %Out-of-equilibrium Kondo effect: Response to pulsed fields
\bibitem{schiller00}
     A. Schiller and S. Hershfield,
      Phys. Rev. B {\bf 62}, R16271 (2000).

\bibitem{Krawiec02}M. Krawiec and K. I. Wysokinski
Phys. Rev. B 66, 165408 (2002)


\bibitem{Rosch01}A. Rosch, J. Kroha and P. W\"{o}lfle:
Phys. Rev. Lett. {\bf 87}, 156802 (2001).


\bibitem{Schiller95}A. Schiller and S. Hershfield:
Phys. Rev. B {\bf 51}, 12896 (1995); {\it ibid}. {\bf 58}, 14978
(1998); K. Majumdar, A. Schiller and S. Hershfield: Phys. Rev. B
{\bf 57}, 2991 (1998): Yu-Wen Lee and Yu-Li Lee: Phys. Rev. B {\bf 65}, 155324 (2002).


 % Out-of-Equilibrium Kondo Effect in Double Quantum, slave p.
 \bibitem{aguado}
     R. Aguado and D. C. Langreth:
    Phys. Rev. Lett. {\bf 85}, 1946 (2000).

\bibitem{slaveColeman} P. Coleman, C. Hooley, Y. Avishai, Y. Goldin,
A. F. Ho: J. Phys.: Condens. Matter 14, L205-L211, (2002).

\bibitem{han}
J. H. Han, preprint cond-mat/0312023.
%A New Mean-Field Theory of the Kondo Resonance at Finite Bias

\bibitem{Konik01}R. M. Konik, H. Saleur and A. Ludwig:
Phys. Rev. Lett. 87, 236801 (2001); Phys. Rev. B 66, 125304 (2002).


\bibitem{spectral}
% Spectral function of the Kondo model in high magnetic fields
A. Rosch, T. A. Costi, J. Paaske, P. W\"olfle, Phys. Rev. B {\bf
68}, 014430 (2003).

\bibitem{Anderson70}P. W. Anderson:
J. Phys. C {\bf 3}, 2436 (1970).

\bibitem{Meir94}N.S. Wingreen, Y. Meir, Phys. Rev. B{\bf 49}, 11040 (1994).


\bibitem{Rammer86}J. Rammer and H. Smith,
Rev. Mod. Phys. {\bf 58}: 323 (1986).

\bibitem{Salmhofer01} % RG for 1pt irred.
M. Salmhofer and C. Honerkamp, Prog. Theo. Physics {\bf 105}, 1
(2001).

\bibitem{SK} H. Schoeller and J. K\"onig, Phys. Rev. Lett. {\bf 84},
3686 (2000); see also M. Keil, Ph.D. thesis, U. G\"ottingen (2002).


\bibitem{gabriel}
G. Sellier: {\em Quantum Impurities in Superconductors and
Nanostructures: Selfconsistent Approximations and Renormalization
Group Analysis}, dissertation, University of  Karls\-ruhe,
%ISBN 3-8322-1337-6
(Shaker, Aachen, 2003).

\bibitem{spinpub} G. Sellier, J. Kroha, and A. Rosch: in preparation.

\bibitem{preparation} A. Rosch, J. Paaske and P. W\"{o}lfle:
in preparation.

\bibitem{Costi94} T. A. Costi and A. C. Hewson,
J. Phys. Cond. Mat {\bf 6}, 2519 (1994).









%\bibitem{Zawadowski67}A. Zawadowski:
%Phys. Rev. {\bf 163}, 341 (1967).

%\bibitem{Appelbaum69}J. Appelbaum and W. F. Brinkman
%Phys. Rev. {\bf 186}: 464 (1969).


%\bibitem{Caroli71}C.Caroli, R. Combescot, P. Nozieres and D. Saint-James:
%J. Phys. C: Solid St. Phys., {\bf 4}, 916 (1971).



%\bibitem{Emery92}V. J. Emery and S. Kivelson:
%Phys. Rev. B {\bf 46}, 10812 (1992).


%\bibitem{Nagaoka65}Y. Nagaoka:
%Phys. Rev. {\bf 138}, A1112 (1965).

%\bibitem{Abrikosov65}A. A. Abrikosov,
%Physics {\bf 2}, 5 (1965), {\it ibid}. {\bf 2}, 61 (1965)

%\bibitem{Suhl65}H. Suhl:
%Phys. Rev. {\bf 138}, A515 (1965);
%H. Suhl and D. Wong,
%Physics {\bf 3}, 17 (1967).

%\bibitem{Zawadowski69}A. Zawadowski and P. Fazekas,
%Z. Physik {\bf 226}: 235 (1969).

%\bibitem{Kadanoff62}L. P. Kadanoff and G. Baym,
%{\it Quantum Statisticsl Mechanics} (Benjamin: New York, 1962).

%\bibitem{Langreth76}D. C. Langreth in
%{\it Linear and Nonlinear Electron Transport in Solids},
%eds. J. T. Devreese and E. Van Doren (Plenum, New York, 1976).

%\bibitem{Yosida65}K. Yosida and A. Okiji
%Prog. Theoret. Phys. {\bf 34}, 505 (1965).

%\bibitem{Brenig70}W. Brenig, J. A. Gonzalez, W. G\"{o}tze and P. W\"{o}lfle:
%Z. Physik {\bf 235}, 52 (1970).

%\bibitem{Okiji70}A. Okiji, A. Kato and H. Shiba:
%Suppl. Prog. Theoret. Phys.{\bf 46}, 182 (1970).

%\bibitem{Andrei83}N. Andrei, K. Furuya and J. H. Lowenstein:
%Rev. Mod. Phys. {\bf 55}, 331 (1983).

%\bibitem{Andrei8082}N. Andrei:
%Phys. Rev. Lett. {\bf 45}, 379 (1980);
%Phys. Lett. {\bf 87A}, 299 (1982).

%\bibitem{Wilson7580}K. G. Wilson:
%Rev. Mod. Phys. {\bf 47}, 773 (1975);
%H. R. Krishna-murthy, J. W. Wilkins and K. G. Wilson:
%Phys. Rev. {\bf B21}, 1003 (1980).

%\bibitem{Magno77}R. Magno and J. G. Adler:
%Phys. Rev. B {\bf 15}, 1744 (1977).



\end{thebibliography}
\end{document}